\pdfoutput=1
\documentclass{JINST}

\title{First investigation of a novel 2D position-sensitive semiconductor detector concept}

\author{D.Bassignana$^a$\thanks{Corresponding
author.}~, M.Fernandez$^b$, R.Jaramillo$^b$, M.Lozano$^a$, F.J.Munoz$^b$, G.Pellegrini$^a$, D.Quirion$^a$, I.Vila$^b$ \\
\llap{$^a$} Centro Nacional de Microelectr\'onica (IMB-CNM,CSIC),\\
Campus Univ. Aut\'onoma de Barcelona\\
08193 Bellaterra, Barcelona (Spain)\\
\llap{$^b$} Instituto de F\'isica de Cantabria IFCA (CSIC,UC),\\
Edificio Juan Jord\'a,
Avenida de los Castros, s/n\\
E-39005 Santander (Spain)\\
E-mail: \email{daniela.bassignana@imb-cnm.csic.es}}

\abstract{This paper presents a first study of the performance of a novel 2D position-sensitive microstrip detector,
where the resistive charge division method was implemented by replacing the metallic electrodes with resistive electrodes made of polycrystalline silicon. A  characterization of two proof-of-concept prototypes with
different values of the electrode resistivity was carried out using a pulsed Near Infra-Red laser. The experimental data were
compared with the electrical simulation of the sensor equivalent circuit coupled to simple electronics readout circuits.
The good agreement between experimental and simulation results establishes the soundness of resistive charge division method
in silicon microstrip sensors and validates the developed simulation as a tool for the optimization of future sensor prototypes. Spatial resolution in the strip length direction depends on the ionizing event position. The average value obtained from the protype analysis is close to 1.2\% of the strip length for a 6 MIP signal. }


\begin{document}

\section{Introduction}

In the last 30 years, semiconductor sensors have been object of great interest as position-sensitive
detectors. Many devices have been developed in order to obtain two coordinates of an ionizing event
using double-sided processing (double-sided microstrip detectors and drift detectors)
or implementing a complex readout system with a large number of electronic channels (pixel
detectors). In this paper a different approach is proposed: a new microstrip single-sided detector that
provides 2D position measurements based on the well established resistive charge division method.

The resistive charge division method has been frequently used in gaseous detectors with resistive
anodes \cite{gas} and studied for silicon pad detectors \cite{pad}, but it has never been implemented
in actual semiconductor microstrip detectors. Radeka, in his seminal paper \cite{Radeka}, formulated for
the first time the basic characteristics of the charge-division concept for resistive electrodes.
One of the main conclusions of this study is the fact that the position resolution -assuming a
readout electronics with optimal shaping time- is determined only by the electrode capacitance and not by
the electrode resistance.

Recently, the use of the charge-division method in very long microstrip sensors -several
tens of centimeters- has been proposed as a possible tracking technology for the
International Linear Collider detector concepts. Along this application
line, the behaviour of a detector equivalent RC network implemented in a PC board was used for
the benchmarking of a SPICE electronic circuit simulation \cite{board}.
The PC board was populated with discrete components with electrical specifications matching the
main electrical parameters of such long microstrip detector DC coupled to the readout electronics. The simulation, supported by the
RC network measurements, confirmed the overall validity of Radeka's formulation on resistive
charge-division.

In this paper, a novel microstrip detector concept is introduced, where the resistive electrodes (light yellow structure in figure \ref{fig:detector} (a) and (b)) are made of a thin layer of highly doped polycrystalline silicon.
This original approach decouples the resistive electrodes from the detector diode structure (determining the charge collection) through a coupling capacitance (gray layer in figure \ref{fig:detector} (b)). In this way, we can easily manufacture AC coupled p+-on-n or n+-on-p detectors, more convenient for the case of high radiation environments. 
The field of application of this device expands beyond the nuclear or particle physics tracking applications, reaching other possible areas as laser-based position sensitive devices, heavy ion and other highly-ionizing particle detector, Compton cameras, medical imaging, etc.

The goal of this investigation is to demonstrate the feasibility of the resistive charge division concept in a
full-fledged microstrip detector. 
We are presenting the first results on the reconstruction of the hit position along the strip
direction using the proof-of-concept prototypes, without any dedicated effort yet to
the conventional reconstruction of the transverse coordinate.
The experimental method used to this purpose was the study of the output signals under a longitudinal scan of a strip, carried out with an infrared laser beam.


\section{Resistive charge division in microstrip detectors}
\subsection{Effect of the resistive readout electrodes on the detector response}

\begin{figure}[hb]
\begin{center}
\begin{tabular}{p{0.55\textwidth}@{\hspace{0.001\textwidth}}p{0.55\textwidth}}
\includegraphics[width=0.3\textwidth]{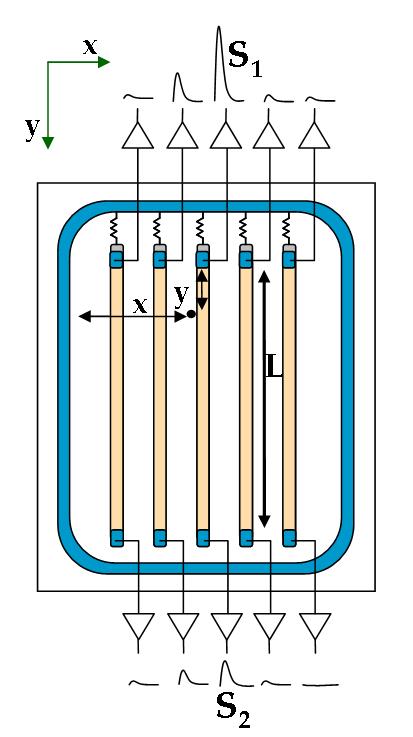} &
\includegraphics[width=0.55\textwidth]{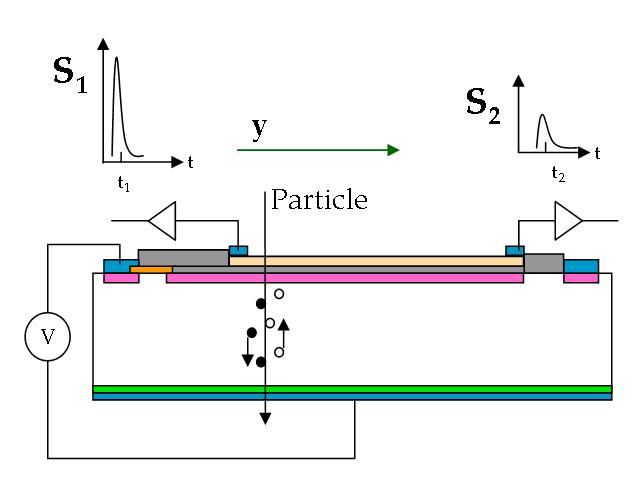} \\
\centerline{(a)}&\centerline{(b)}\\

\end{tabular}
\end{center}
\vspace{-0.7 cm}
\caption{\footnotesize{Schematic top view of the novel detector (a) and lateral cross-section of the central strip (b)(not to scale). It is possible to distinguish the aluminum elements in blue and
the resistive electrodes on the strips (light yellow regions). The aluminum pads are connected each one to a channel of
the read-out electronics (two for each strip). When an ionizing particle crosses the detector, different signals ($S_1$ and $S_2$) are read by the opposite electronic channels. The X coordinate of the event can be reconstructed
using the center of gravity method, whereas the Y coordinate is reconstructed comparing the signal amplitudes at the ends of  strips.}}
\label{fig:detector}
\end{figure}

In a conventional microstrip detector the metal contacts of the strips extend over almost all the length of the implants and are connected each one to a read-out channel. When an ionizing particle crosses the detector, the propagation of the induced signal along the coupling electrode does not suffer significant attenuation, i.e., the signal amplitude does not depend on the particle impinging point along the electrode direction. When using, instead of metal alloy, a resistive coupling electrode equipped with metal contacts at its ends, the signal undergoes significant attenuation during its propagation towards the electronics contacts. The longer the propagation length, the larger the signal attenuation. 
In this way, a conventionally manufactured single-sided microstrip sensor can provide the two-dimensional coordinates of the particle impinging point; the transverse coordinate derived from the usual electrode segmentation \cite{Tur} and the longitudinal coordinate determined by relating signal amplitude at both ends of the electrode.

The resistive electrode represents a diffusive RC line, in which a current pulse undergoes not only an amplitude attenuation but also an increase of the rise time the further it travels. Using readout electronics characterized by a short -compared to the RC constant of the line- shaping time, this translates into a non constant signal ballistic deficit. Increasing the electrode resistance also increases the readout serial noise contribution. Both the ballistic deficit and the serial noise can be reduced increasing the shaper peaking time; however, a longer peaking time increases the parallel readout noise contribution. In reference \cite{Radeka}, Radeka derived the optimal peaking time for a resistive charge division configuration, under the assumption of high electrode resistance compared to the amplifier impedance and long amplifier peaking time compared to input signal rise times. He concluded that the position resolution achieved with the resistive charge division method should be independent of the electrode resistance, depending only on detector capacitance and signal amplitude.

Under these assumptions, we expect a linear dependence between the longitudinal coordinate of the particle position and the fractional signal amplitude read from one side of the strip. The actual functional form (following figure \ref{fig:detector} notation with $A_1$ and $A_2$ the amplitudes of $S_1$ and $S_2$ respectively) is given by equation \ref{equ1}:

\begin{equation}\label{equ1}
y=L\times\frac{A_2}{A_1+A_2}
\end{equation}

\subsection{Proof-of-concept prototype: specifications and expected performance}

\begin{figure}[hbt]
\begin{center}
\includegraphics[height=0.35\textheight,angle = 0]{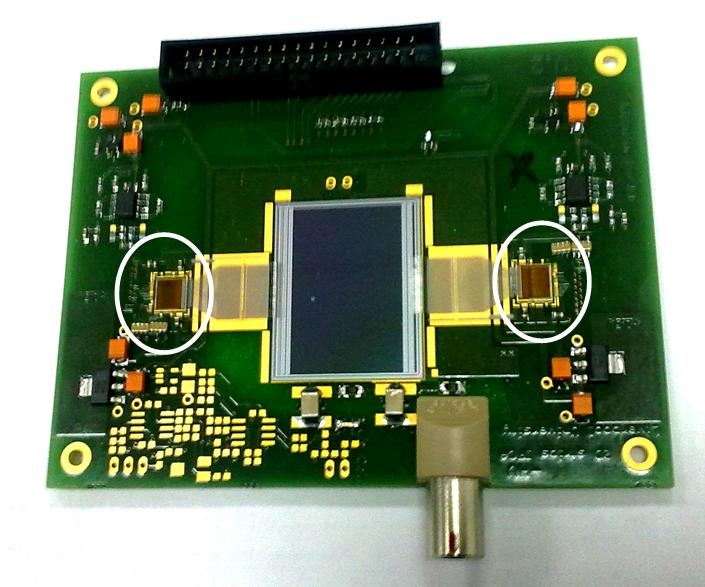}
\caption{\footnotesize{Picture of one of the detectors mounted in the PCB sensor carrier. The two Beetle chips are indicated by the white circles. Each one is connected to one side of 128 consecutive strips of the detector in order to provide double-sided readout.}}
\label{fig:foto}
\end{center}
\end{figure}

The resistive charge-division proof-of-concept prototypes studied here are AC coupled microstrip detectors with the upper electrodes of the coupling capacitor made of polycrystalline silicon. Figure \ref{fig:foto} shows a picture of the actual prototype attached to the sensor carrier. Two prototypes with different electrode resistivities were fabricated at the IMB-CNM clean room facilities in Barcelona \cite{cnm} using the conventional technology for single-sided p+-on-n, AC coupled, silicon microstrip detectors. A standard reference sensor and several electrical test structures \cite{TSVienna} were included in each wafer of the fabrication run that allowed a more direct measurement of the electrical parameters of the new sensors.

Each detector consists of 384 p$^+$ strips ($20~\mu$m wide) with a pitch of $80~\mu$m on a (285$\pm$15)~$\mu$m thick n-type substrate. The resistive electrodes have a total length of 2~cm with linear resistance $R\slash l$=2.8~$\Omega \slash \mu$m for one of the devices and $R\slash l$=12.2~$\Omega \slash \mu$m for the other.
The detectors have been electrically characterized in the IMB-CNM laboratories with the use of a  Cascade Microtech probe station, two Keithley 2410 Source/Meters and an Agilent 4284A LCR Meter. The results are consistent with the ones of the standard microstrip detectors.
The measured values are listed in table~\ref{tab:elcar}.

\begin{table}[h]
\begin{center}
\caption{\footnotesize{Electrical characterization: measured values of the polycrystalline silicon electrode resistance, depletion voltage, breakdown
voltage, bias resistance, interstrip resistance, interstrip capacitance and coupling capacitance.}}
\vspace{0.5 cm}
\begin{tabular}{|c||c|c|c|c|c|c|}
\hline
electrode resistance & $V_{dep}$~[V] & $V_{bd}$~[V] & $R_{bias}$~[M$\Omega$] & $R_{int}$ & $C_{int}$~[pF/cm]& $C_{AC}$~[pF/cm] \\
\hline\hline
2.8~$\Omega$/$\mu$m & 20 & \textgreater 300 & 4 & \textgreater G$\Omega$ & 0.46 & 182 \\
\hline
12.2 $\Omega$/$\mu$m & 20 & \textgreater 400 & 2.41 & \textgreater G$\Omega$ & 0.46 & 182 \\
\hline
\end{tabular}
\label{tab:elcar}
\end{center}

\end{table}

\begin{figure}[hbt]
\begin{center}
\includegraphics[height=0.35\textheight,angle = 0]{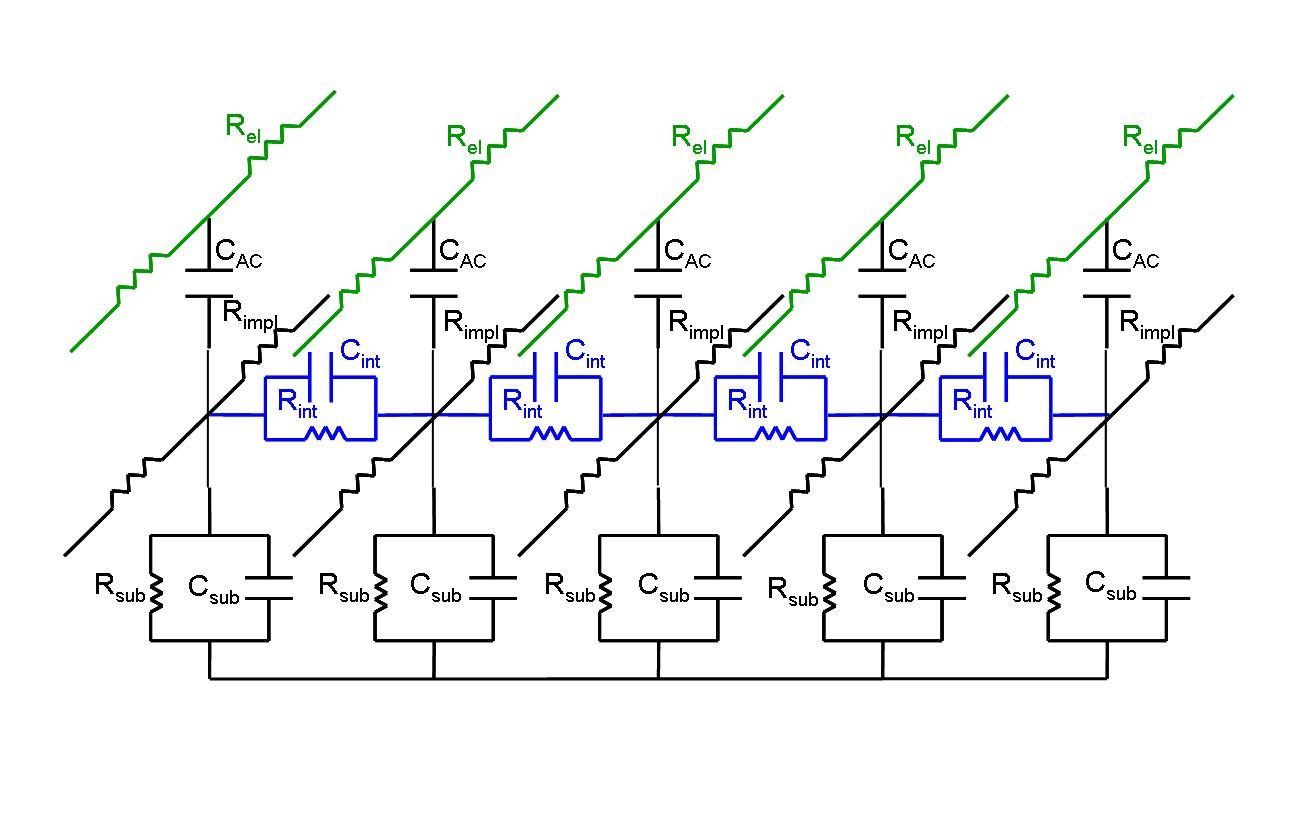}
\caption{\footnotesize{Schematics of one of the 80 cells used to model the detector. Each one
represents a portion (250~$\mu$m long) of five consecutive strips including the main electrical parameters like the coupling
capacitance ($C_{AC}$), the substrate resistance and capacitance ($R_{sub}$ and $C_{sub}$), the p$^{+}$implant
resistance ($R_{impl}$) and the resistance of the resistive electrode ($R_{el}$). In the simulation a current pulse has
been induced at different nodes along the central strip implant.}}
\label{fig:simulation_cell}
\end{center}
\end{figure}

\begin{figure}[hbt]
\begin{center}
\includegraphics[height=0.20\textheight,angle = 0]{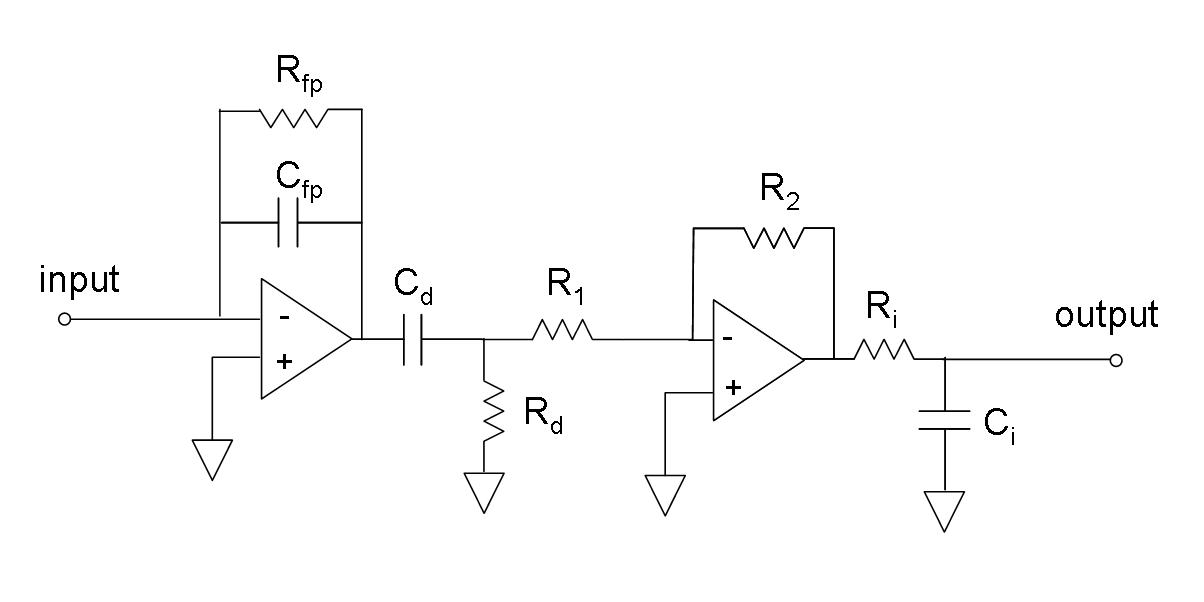}
\caption{\footnotesize{Schematics of the front-end electronics modeled for the simulation.}}
\label{fig:electronics}
\end{center}
\end{figure}

For our study, no dedicated analog signal processing electronics was built and therefore the front-end filtering of
the signal was not optimized according to Radeka's conclusions. As it will be explained in more detail in the following section, we have used the ALIBAVA DAQ system  \cite{alibava} developed within the framework of the CERN RD50 collaboration. The analog front-end of the ALIBAVA system is based on the Beetle chip  \cite{beetle} used for the microstrip sensor readout of the silicon tracking subsystem of the LHCb experiment at LHC;
consequently, the analog front-end shaper peaking time is set around 25~ns.

We have developed a SPICE-like model of each prototype and of the readout electronics in order to clarify the possible effect of the non-optimal shaping time on the linearity of the equation \ref{equ1}, studying the response of the detector to a simulated current pulse injected at different points along the strip length.

Starting from the work presented in reference~\cite{spice} we developed the model of our detectors
built with standard components from the AnalogLib library of Virtuoso Spectre by Cadence \cite{Spectre}.
A portion of the detector including five consecutive strips is modeled by a periodic structure composed of 80 cells,
each one corresponding to a transverse section ($250~\mu$m long) of the strips. The unit cell is a complex
chain of capacitances and resistors representing the main electrical characteristics of
the device as the substrate resistance and capacitance ($R_{sub}$ and $C_{sub}$), the interstrip resistance
and capacitance ($R_{int}$ and $C_{int}$), the p$^{+}$implant resistance ($R_{impl}$), the coupling capacitance
($C_{AC}$) and the resistance of the resistive upper electrode ($R_{el}$). In figure~\ref{fig:simulation_cell}
the schematic of the unit cell is shown. The values of the circuital elements have been determined from the
ones measured during the electrical characterization of the detectors in full depletion ($V_{bias}$=40V) and of the test structures. These values
are listed in table~\ref{tab:sim_par}.

\begin{table}[h]
\begin{center}
\caption{\footnotesize{List of the values of the model parameters. Detector on the left, readout electronics on the right.}}
\vspace{0.5 cm}
\begin{tabular}{|c|c||c|c|}
\hline
$R_{el}$ & $350~\Omega$ or $1525~\Omega$ & $R_{fp}$ & 300~M$\Omega$ \\
\hline
$R_{impl}$ & $718~\Omega$ & $C_{fp}$ & 1pF \\
\hline
$C_{AC}$ & 4.7~pF & $C_{d}$ = $C_{i}$ & 25pF \\
\hline
$C_{sub}$ & 8.6~fF & $R_{d}$ = $R_{i}$ & $1~k\Omega$ \\
\hline
$R_{sub}$ & $20000~G\Omega$ & $R_{1}$ & $1~k\Omega$ \\
\hline
$R_{int}$ & $15~G\Omega$ & $R_{2}$ & $1~k\Omega$ \\
\hline
$C_{int}$ & 11.5~fF & $R_{t}$ & $1~M\Omega$ \\
\hline
\end{tabular}
\label{tab:sim_par}
\end{center}

\end{table}

\begin{figure}[hbt]
\begin{center}
\includegraphics[height=0.30\textheight,angle = 0]{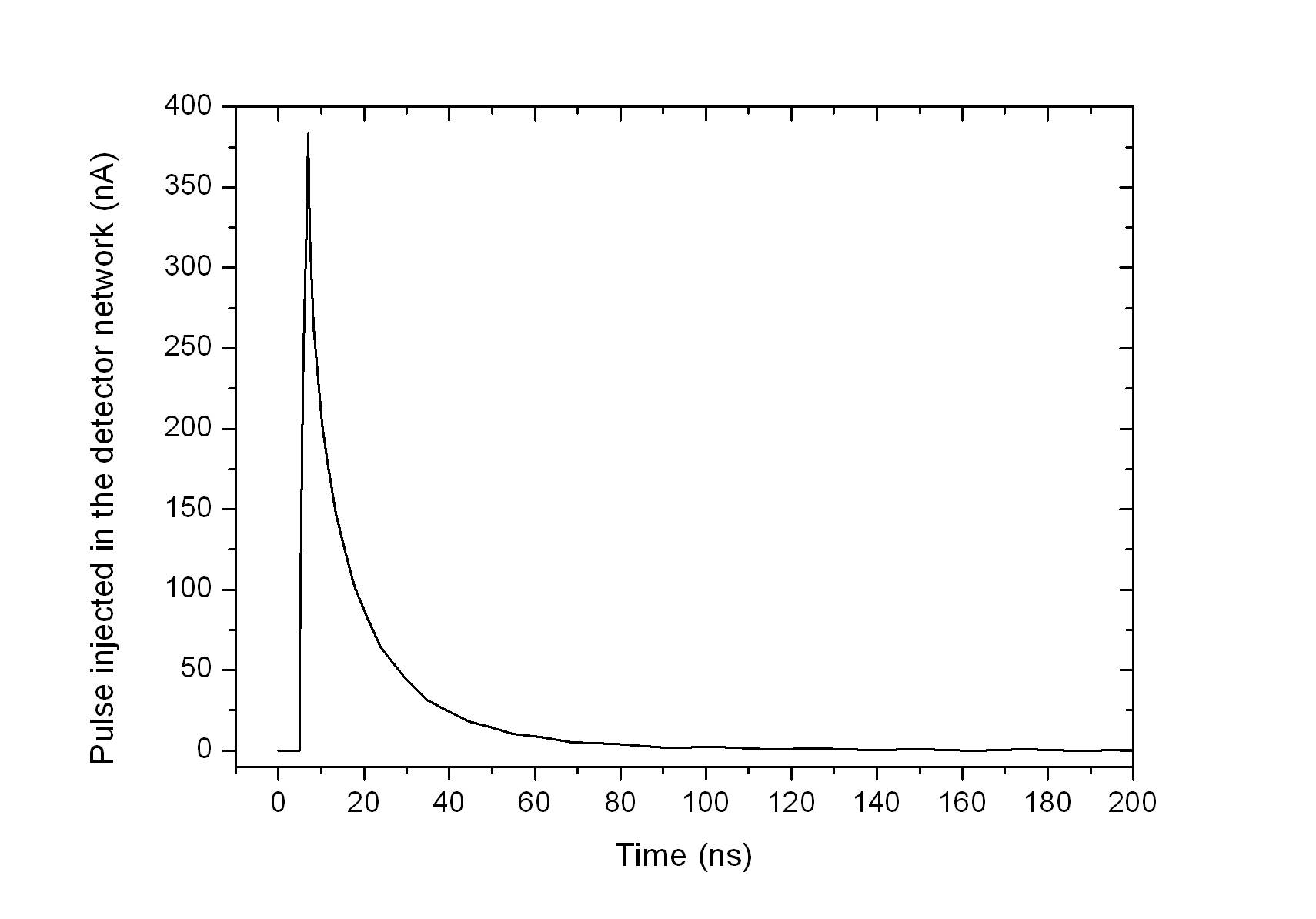}
\caption{\footnotesize{Simulated input signal.}}
\label{fig:pulse}
\end{center}
\end{figure}

The shape of the injected current is shown in figure~\ref{fig:pulse}. It is characterized by a rise time of 2~ns and total integrated charge
around 4~fC. The rise time of the diode laser we have used for our study is
around 2~ns (measured with a high bandwidth photodiode), similar to the simulated one.
The model of the read-out electronics connected to the ends of each strip consists in a generic charge
sensitive preamplifier followed by a CR-RC filter, whose peaking time matches that of the Beetle chip. The front-end schematics
is shown in figure \ref{fig:electronics} and the parameter values are listed in table \ref{tab:sim_par}.

\begin{figure}[!ht]
\begin{center}
\begin{tabular}{p{0.55\textwidth}@{\hspace{0.001\textwidth}}p{0.55\textwidth}}
\includegraphics[width=0.55\textwidth]{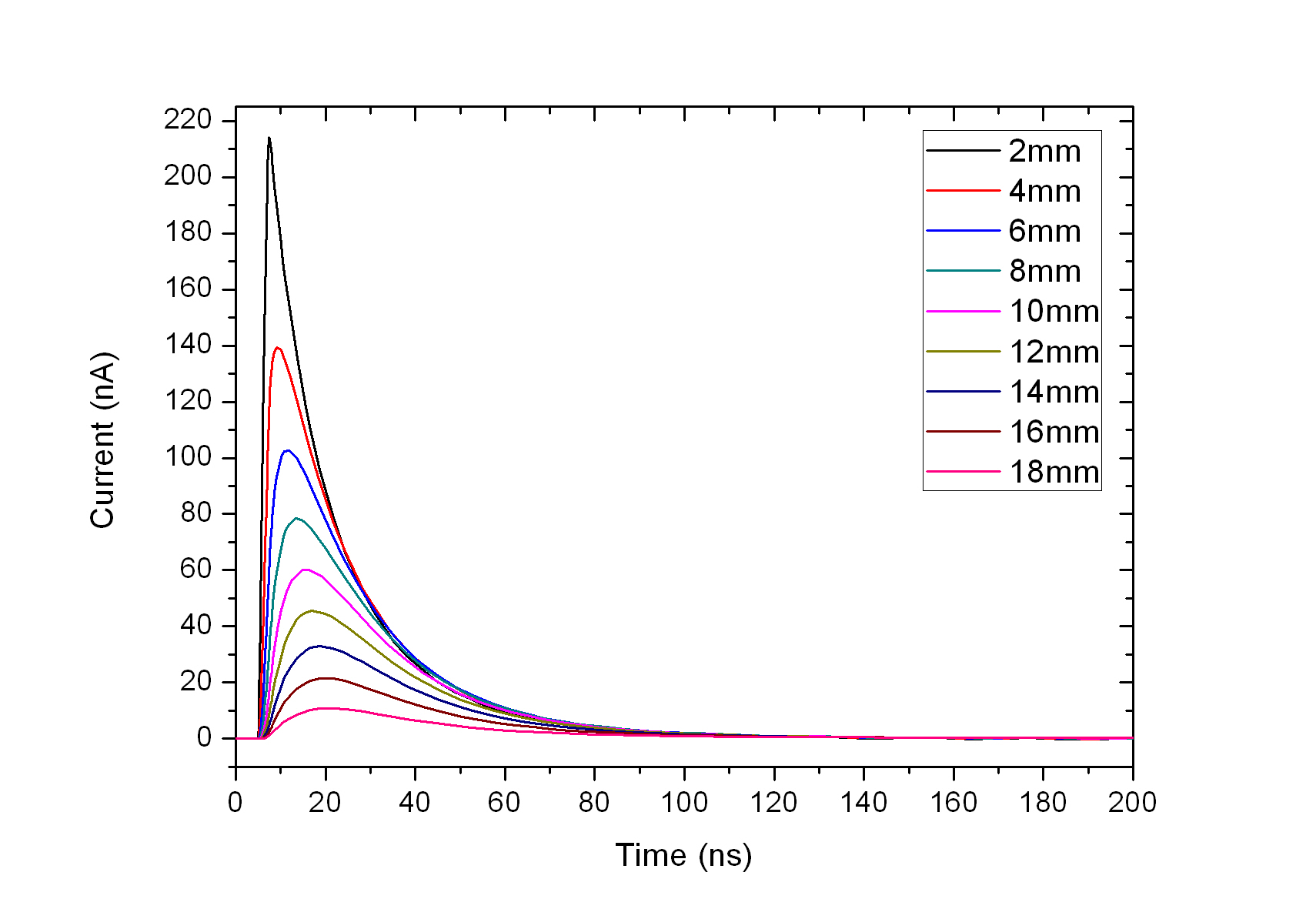} &
\includegraphics[width=0.55\textwidth]{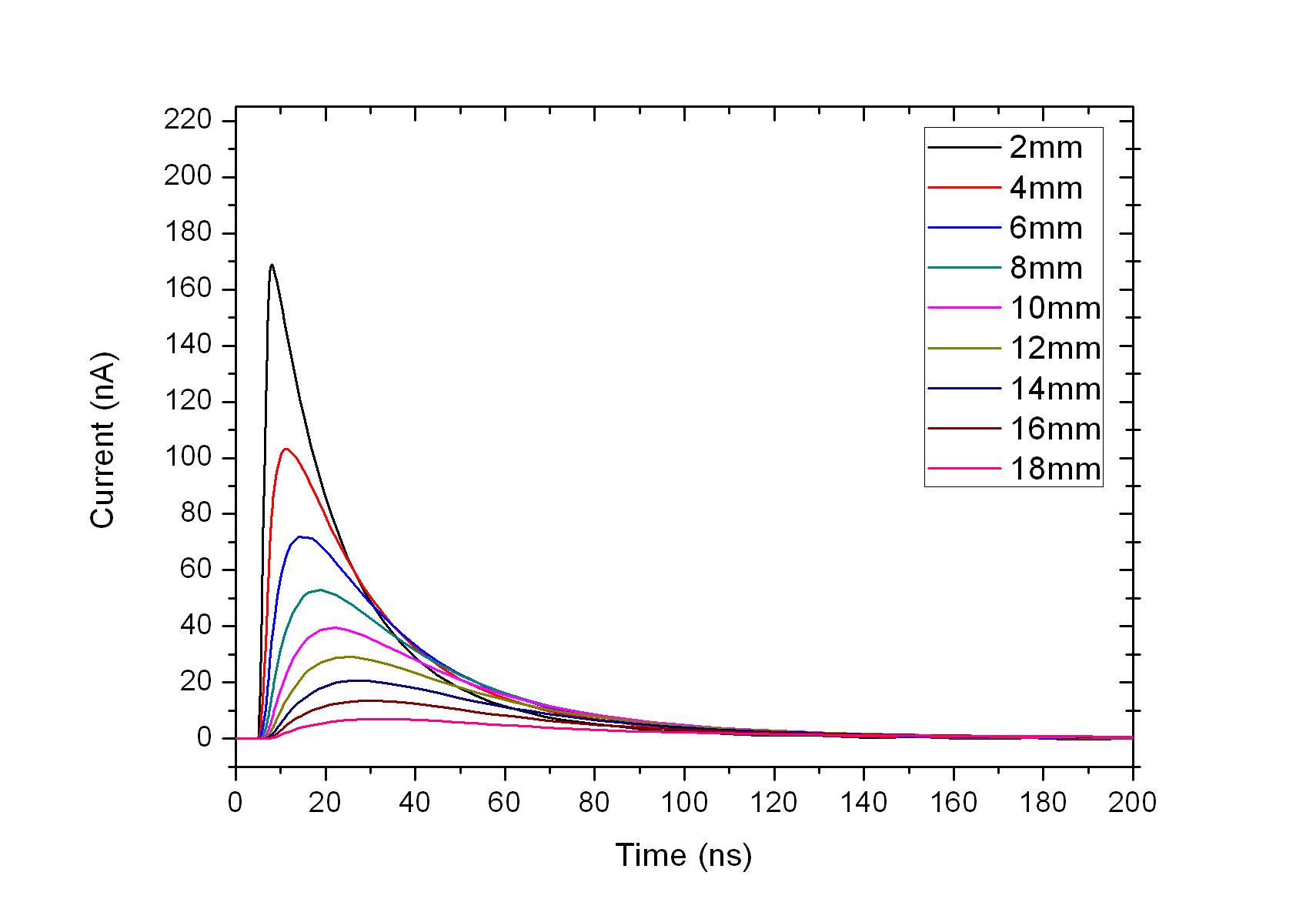} \\
\centerline{(a)}&\centerline{(b)}\\

\end{tabular}
\end{center}
\vspace{-0.7 cm}
\caption{\footnotesize{Attenuation of the signal read at the entrance of the amplifier connected to the strip end in the corresponding position 0~mm. The results are shown for different positions of the pulse generator along the strip for $R\slash l$=2.8~$\Omega \slash \mu$m (a) and $R\slash l$=12.2~$\Omega \slash \mu$m (b).}}
\label{fig:att}
\end{figure}

The signal generator was connected to different points along the implant of the central strip with a step of 2 mm and the shapes of the current pulses, propagated to the entrance of the opposite charge-sensitive preamplifiers, have been recorded as well as the output response of the shapers. The current pulse read at the entrance of the preamplifier connected to the first cell of the strip (corresponding to the position 0~mm) is shown in figure~\ref{fig:att} for each injection point and for both prototypes.
As expected, the simulation confirms that the further the pulse travels the more its amplitude decreases and its rise time increases and stronger effects can be observed for higher values of the electrode resistance.

\begin{figure}[!ht]
\begin{center}
\begin{tabular}{p{0.55\textwidth}@{\hspace{0.001\textwidth}}p{0.55\textwidth}}
\includegraphics[width=0.55\textwidth]{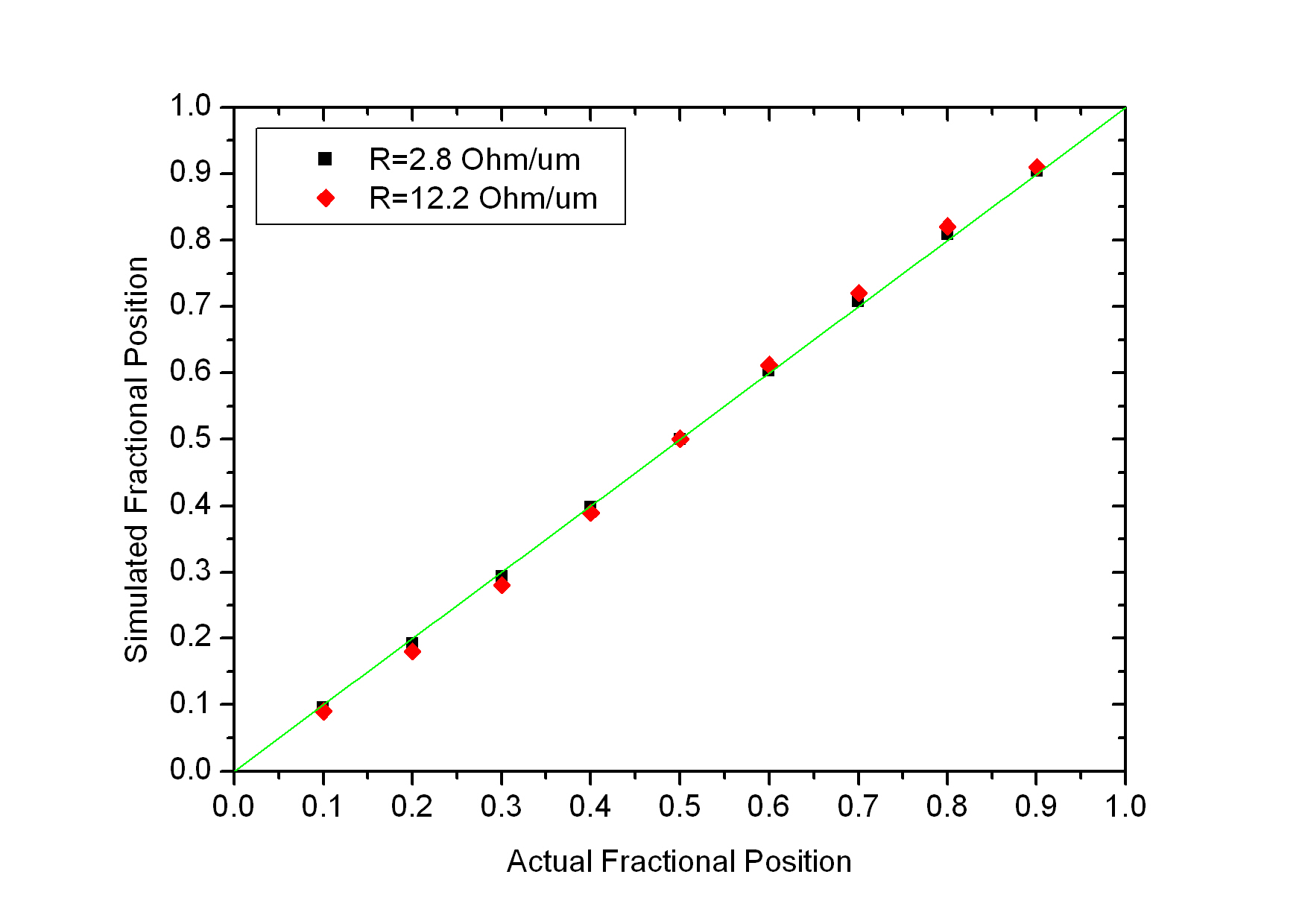} &
\includegraphics[width=0.55\textwidth]{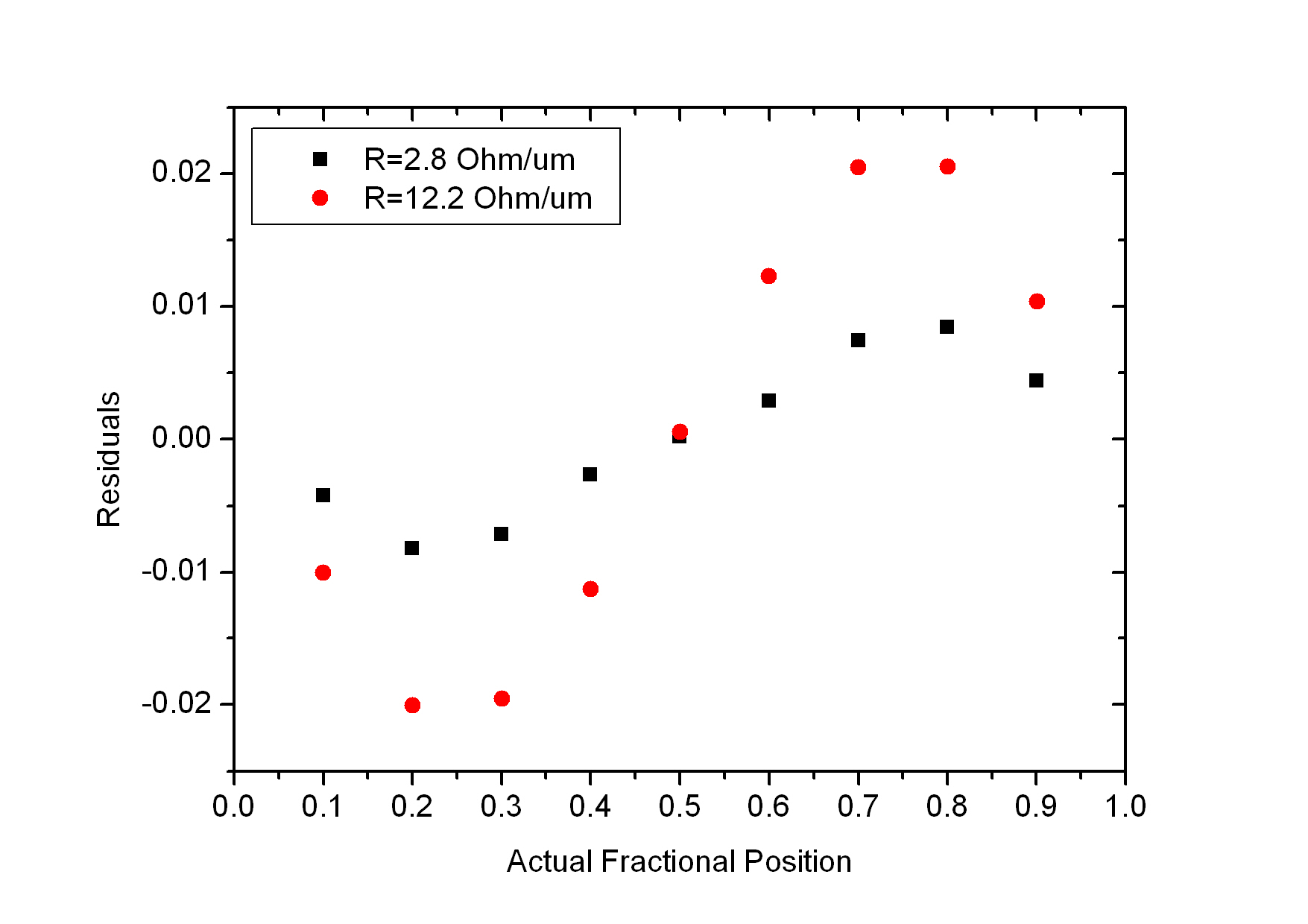} \\
\centerline{(a)}&\centerline{(b)}\\
\end{tabular}
\end{center}
\vspace{-0.7 cm}
\caption{\footnotesize{Simulated fractional position against the actual one for both values of the electrode linear resistance: $R\slash l$=2.8~$\Omega \slash \mu$m and $R\slash l$=12.2~$\Omega \slash \mu$m (a). The values have been compared to the linear prediction (green line). The residual plot (b) shows that for a peaking time of the analog front-end electronics equal to 25~ns, the linearity of the response of the detector suffers higher degradation for higher values of the electrode resistance.}}
\label{fig:fit_sim}
\end{figure}

The dependence of the response linearity on the resistance is clearly seen in figure~\ref{fig:fit_sim}.
According to equation \ref{equ1}, with $A_1$ and $A_2$ the amplitudes of the signal read at the output of the front-end electronics connected to the first and the last cells of the central strip respectively, the derived
fractional position ($A_2/(A_1+A_2)$) versus the injection point ($y/L$) of current signal is shown. 
The data related to the more resistive prototype reveal a clear separation (larger residuals spread) from the ideal values due to the ballistic deficit whose effect increases with the distance covered by the signal from the point of generation.
It is worth to note also that in the case of the low resistivity prototype, even for the more attenuated pulse, the rise time is around 10~ns, still only at
40\% of the Beetle peaking time. Therefore the effect of the ballistic deficit in this sensor is highly suppressed.


\section{Laser characterization of proof-of-concept prototypes}

\subsection{Experimental setup}

Each sensor was mounted in a dedicated PCB sensor carrier and read out using an ALIBAVA DAQ system. The ALIBAVA is a DAQ system for the readout of microstrip sensors based on the Beetle analog readout ASIC. The Beetle integrates 128 pipelined channels with low-noise charge-sensitive preamplifiers and shapers with a peaking time of about 25~ns. Each detector board has two Beetle chips, each one bonded to one side of 128 strips of the sensor like in figure\ref{fig:foto}.

\begin{figure}[hbt]
\begin{center}
\includegraphics[height=0.35\textheight,angle = 0]{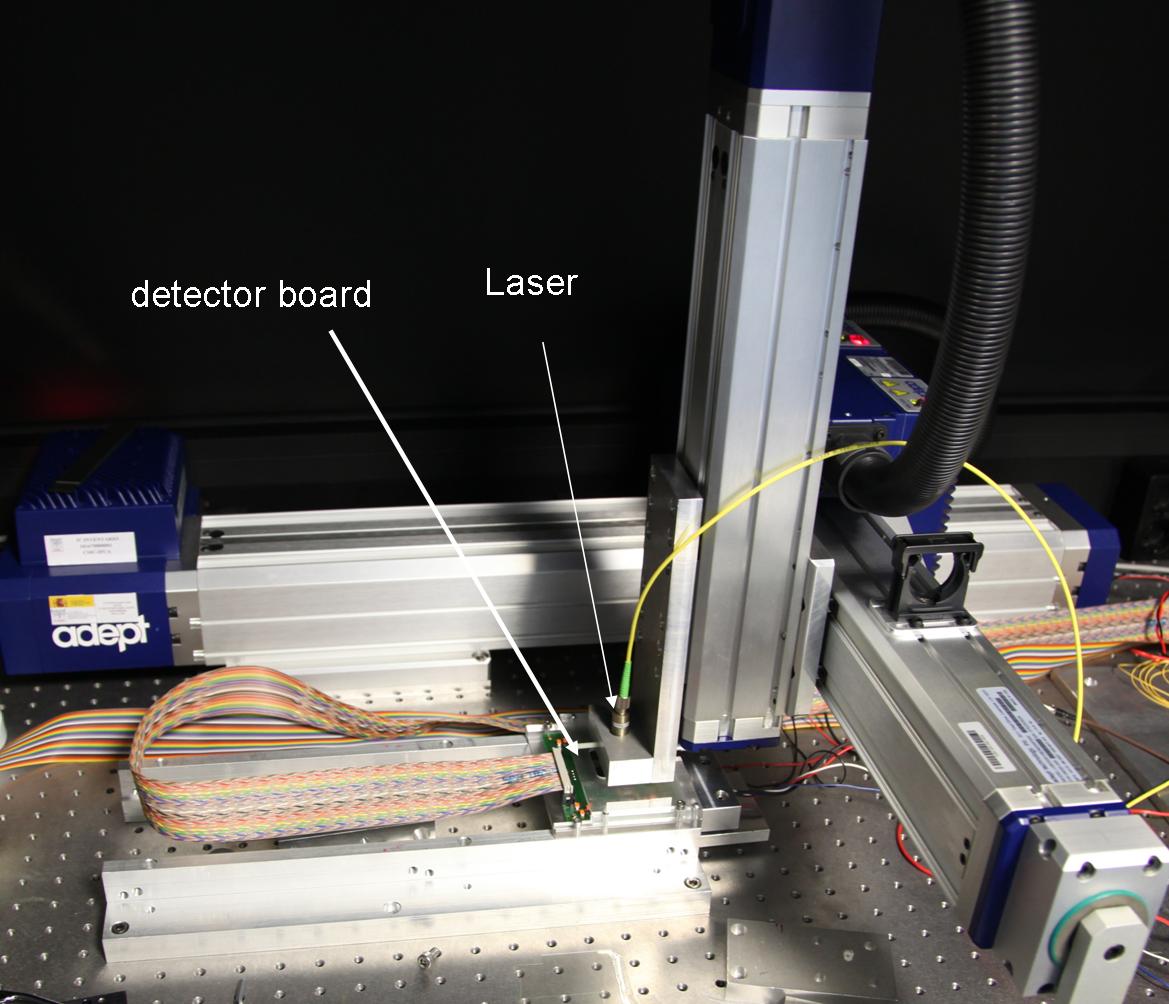}
\caption{\footnotesize{Experimental setup. The micro focusing optical head mounted on the 3D axes stage is placed a few millimeters upon the detector board.}}
\label{fig:laser_setup}
\end{center}
\end{figure}

The characterization test-stand allows for the precise injection of laser pulses along the microstrip direction (see figure \ref{fig:laser_setup}). We used a pulsed distributed-feedback diode laser driven in a constant optical power mode and thermally stabilized. The laser output is coupled to a monomode optical fiber which feeds an inline fiber optic splitter: the first splitter output fiber is connected to a large bandwidth (2~Ghz) reference photodiode whose output signal is recorded in a digital scope to monitor the laser pulse trace; the second splitter output fiber feeds a microfocusing optical head illuminating the sensor. The microfocusing optical head was moved by a 3D axes stage with a displacement accuracy better than  $10~\mu$m for all the axes. The laser is focused in such a way that the beam waist is at the sensor front plane; the beam intensity profile at the beam waist is a Gaussian with a sigma of $5~\mu$m. The laser wavelength is centered at 1060~nm and the laser rise time, as measured by the reference photodiode, is 2~ns.

Comparing the signal amplitudes obtained with the laser pulses and the signal amplitudes obtained using a $^{90}$Sr beta source, we estimated that the charge ionized by the laser pulse used during the sensor characterization is roughly equivalent to six times the most probable charge ionized by a minimum  ionizing particle (MIP). 
The pulse optical power was adjusted using inline optical fiber attenuators and by tweaking the working parameters of the laser driver.



\subsection{Results and discussion}

For each detector, we performed a longitudinal scan moving the focused beam spot along the midline of a polysilicon electrode -contrary to aluminum, polycrystalline silicon is transparent to IR light. We scanned the whole electrode length (20 mm) with a scanning step of 2~mm, reconstructing, for each position, the pulse shape at the output of the front-end electronics shaper stage.

\begin{figure}[!hb]
\begin{center}
\begin{tabular}{p{0.55\textwidth}@{\hspace{0.001\textwidth}}p{0.55\textwidth}}
\includegraphics[width=0.55\textwidth]{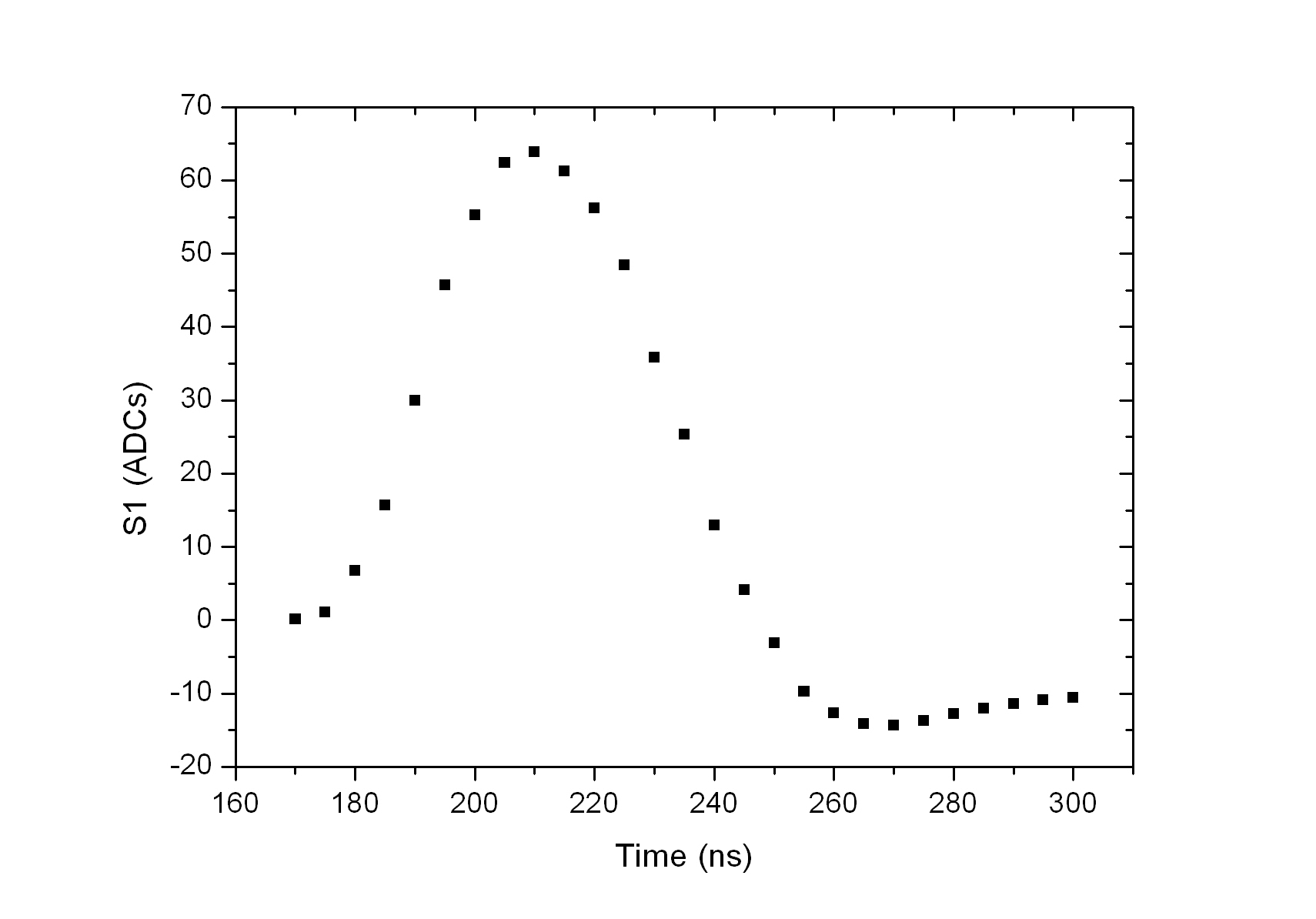} &
\includegraphics[width=0.5\textwidth]{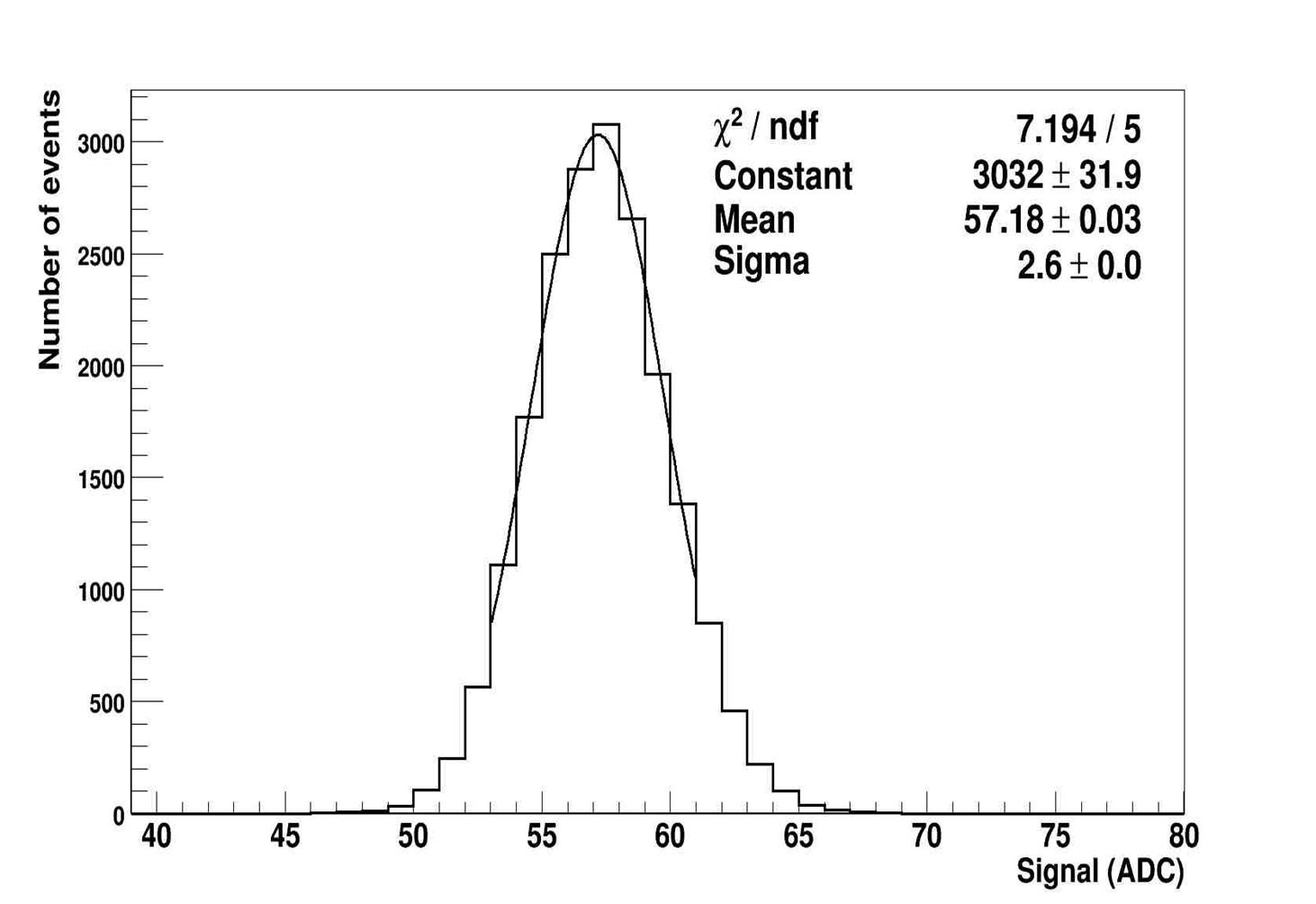} \\
\centerline{(a)}&\centerline{(b)}\\
\end{tabular}
\end{center}
\vspace{-0.7 cm}
\caption{\footnotesize{Reconstruction of the pulse shape of the signal S$_1$ when the laser is 6 mm far from the contact pad along the scanned strip. The electrode linear resistance is  $R\slash l$=2.8~$\Omega \slash \mu$m} (a). For each time delay 20000 events have been recorded and the mean value of the signal has been extrapolated by fitting a Gaussian function to the data. As an example of this procedure the particular case of time delay = 200~ns is shown on the right (b).}
\label{fig:shape}
\end{figure}

The ALIBAVA DAQ system does not allow to record the whole shape of the analog signal. On the other hand it allows to reconstruct it thanks to a particular feature that permits to change the value of the delay between the trigger time (synchronous with the laser pulse) and the acquisition time (specifying the instant at which the shaper output is sampled) \cite{alibava}. Setting different delays in steps of 5~ns from 170 to 300~ns (see figure \ref{fig:shape} (a)), we recorded 20000 events for each time delay and we found the amplitude of their distributions by fitting a Gaussian function (figure \ref{fig:shape} (b)) obtaining a strong suppression of the statistical error. 
The amplitudes of the reconstructed pulses have been accurately extrapolated by fitting a Gaussian function to the peak region. These values have been used for the calculation of the fractional position defined by equation \ref{equ1}.

\begin{figure}[!ht]
\begin{center}
\begin{tabular}{p{0.55\textwidth}@{\hspace{0.001\textwidth}}p{0.55\textwidth}}
\includegraphics[width=0.55\textwidth]{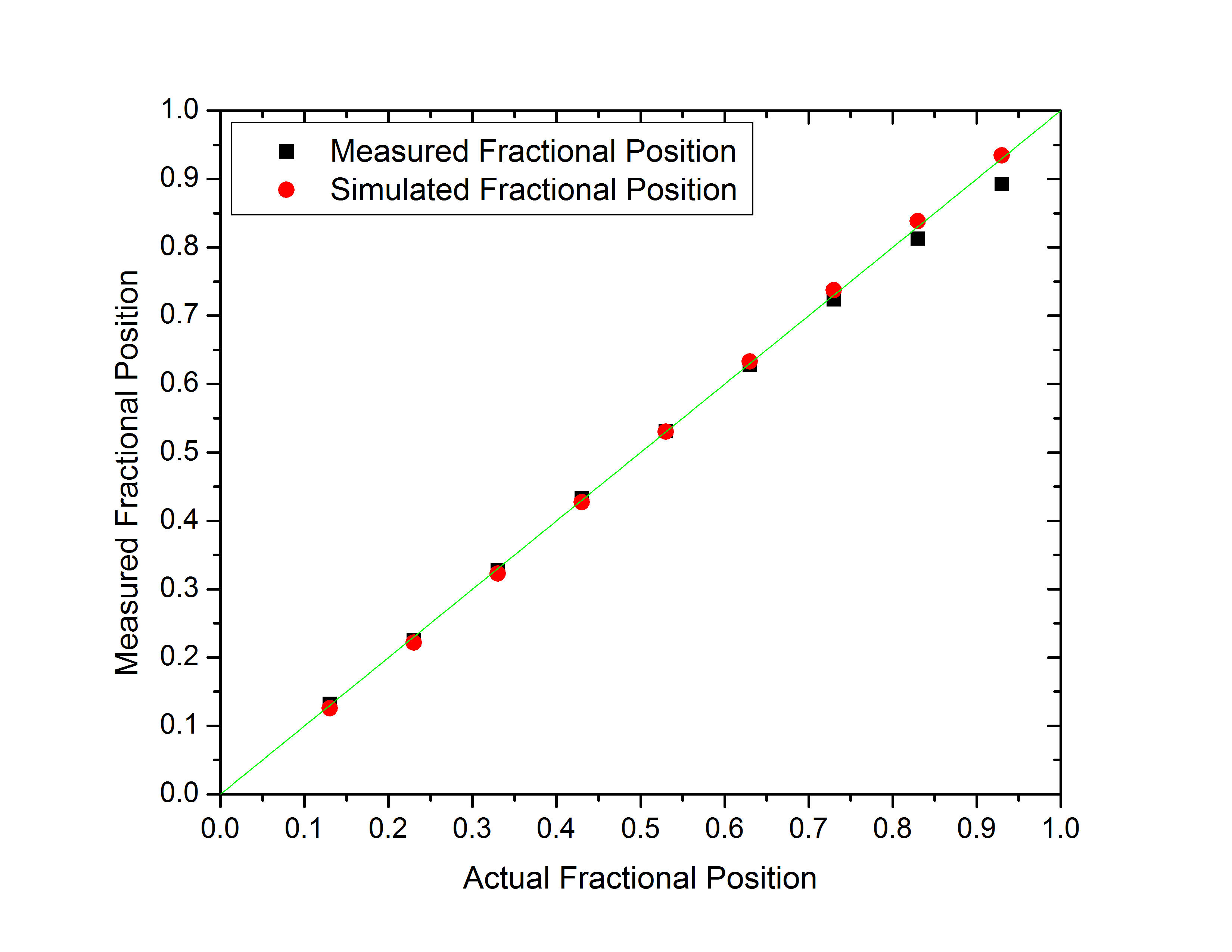} &
\includegraphics[width=0.55\textwidth]{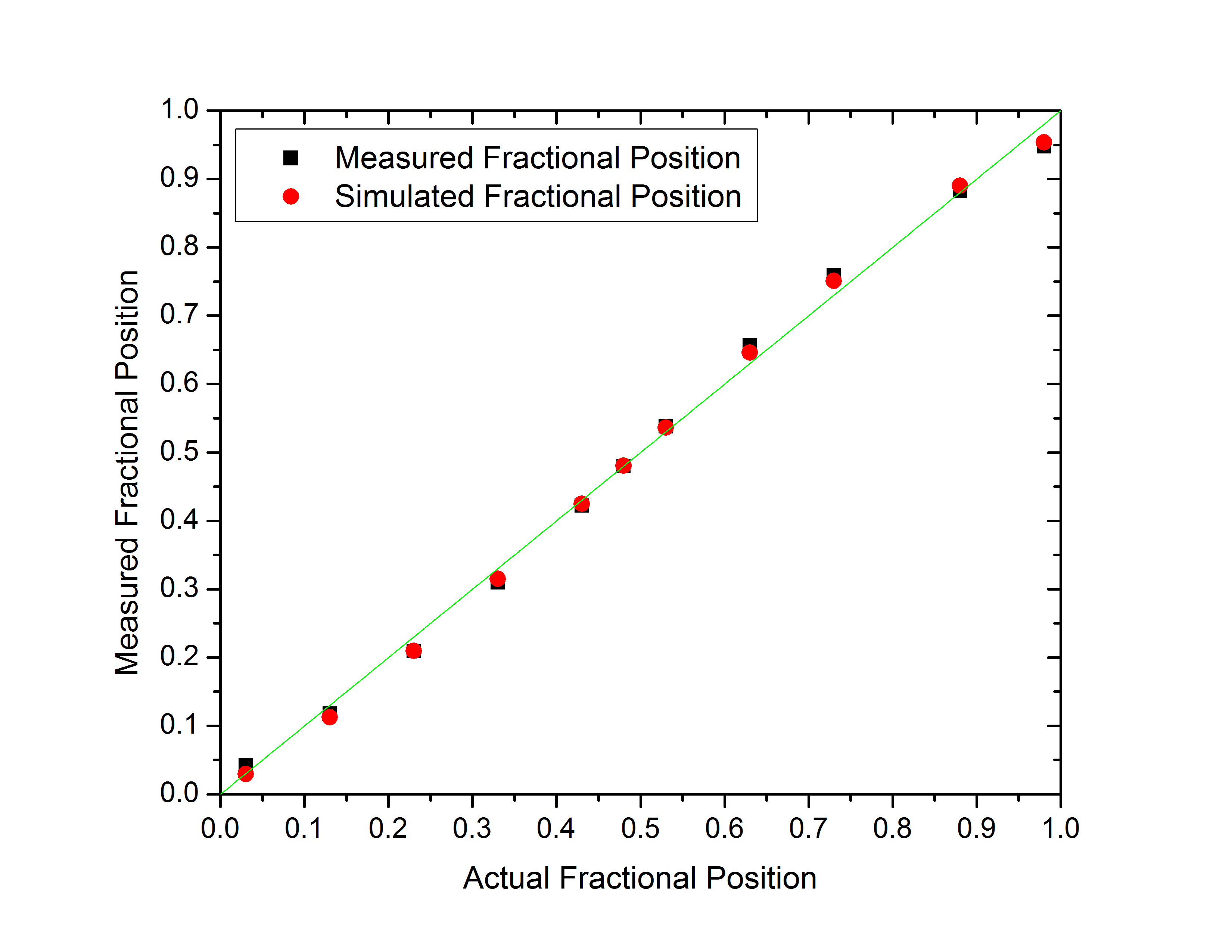} \\
\centerline{(a)}&\centerline{(b)}\\

\end{tabular}
\end{center}
\vspace{-0.7 cm}
\caption{\footnotesize{Experimental results compared with the simulation and the ideal case (green line) for both values of the electrode resistance: $R\slash  l$ = 2.8~$\Omega \slash \mu$m (a) and $R\slash  l$ = 12.2~$\Omega \slash \mu$m (b).}}
\label{fig:sim_exp}
\end{figure}

Figure \ref{fig:sim_exp} shows, for both sensors, the measured fractional position of the laser spot against the
position given by the displacement of the micrometric stage. 
The comparison with the ideal linear behavior given by equation \ref{equ1}
is shown as well as the comparison with the simulation data. We observe the degradation of the linearity of the detector response due to the systematic error introduced by the non-constant ballistic deficit: the higher the value of the electrode resistivity, the deeper the discrepancy between the data and the expected values. 
At this stage, before any noise considerations (it has been considerably averaged out in the experimental analysis), the simulation data and the experimental data show a similar systematic behavior. The good agreement between these results confirms that the electrical simulation reproduces properly the systematic errors due to a non-optimal shaping time and as a consecuence, this study validates it as an important design tool for sensor optimization.
Actually, in order to meet with different requirements on the strip geometry and on the shaping time of the readout electronics,  it is possible to tune the electrode resistivity without affecting the charge collection behaviour of the sensor, as the resistive electrodes are decoupled from the diode structure of the sensor.
\begin{figure}[hbt]
\begin{center}
\includegraphics[height=0.35\textheight,angle = 0]{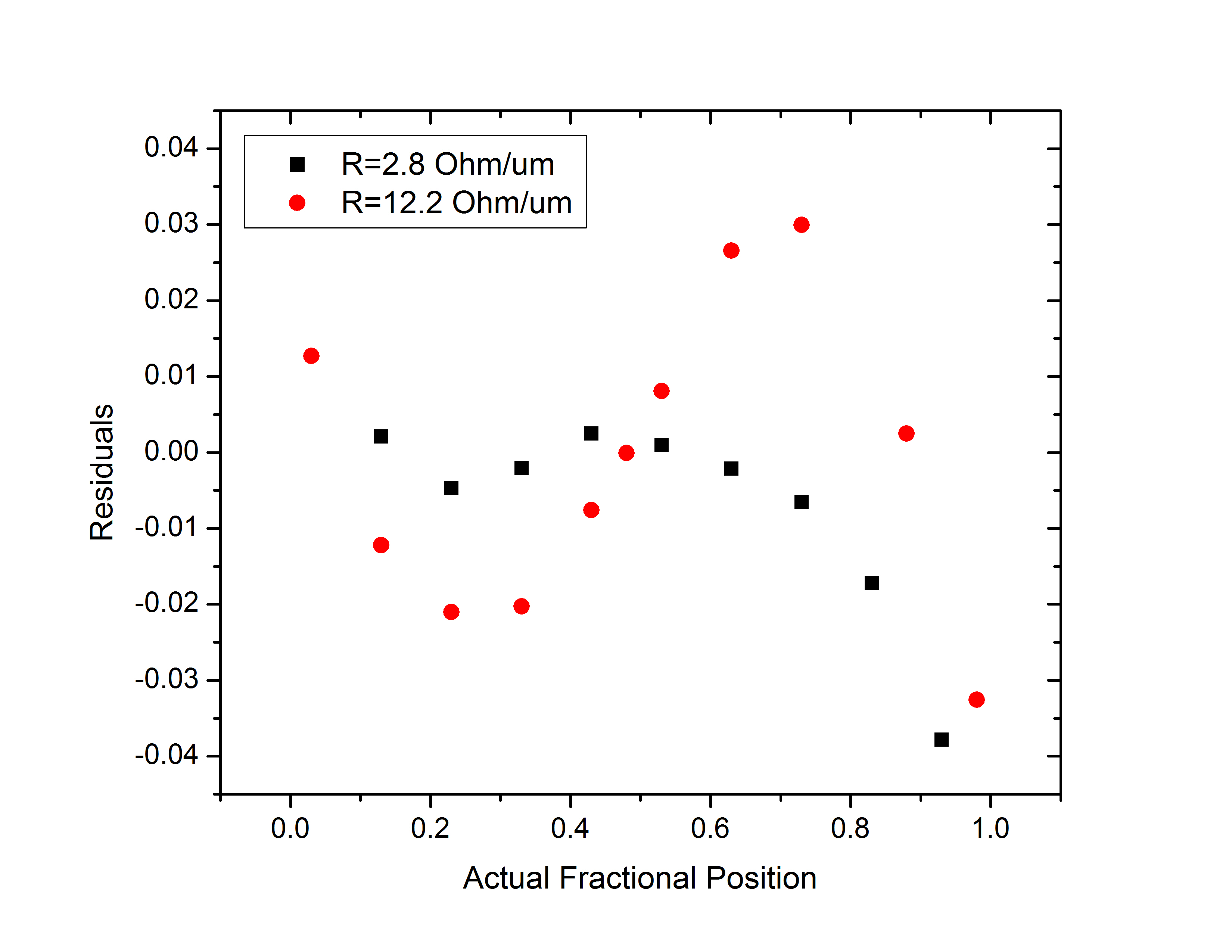}

\caption{\footnotesize{Residuals of the experimental results.}}

\label{fig:res_exp}
\end{center}
\end{figure}
We can also remark how the residuals of the low resistive electrode sensor (see figure \ref{fig:res_exp}) increase for larger values of the fractional position: this effect was caused by the existence of a slight misalignment between the stage scanning direction and the electrode.

Considering the average noise fluctuation registered by the ALIBAVA system for each sensor, it is possible to estimate the fractional position error for 6 MIP signals, using equation:

\begin{equation}\label{equ2}
 \sigma=\frac{A_1A_2}{\left(A_1+A_2\right)^2}\sqrt{\left(\frac{\sigma_{A_1}}{A_1}\right)^2+\left(\frac{\sigma_{A_2}}{A_2}\right)^2-2\rho \left(\frac{\sigma_{A_1}}{A_1} \frac{\sigma_{A_2}}{A_2}\right)},
\end{equation}

with $\sigma_{A_1}$ and $\sigma_{A_2}$ the noise fluctuations of $A_1$ and $A_2$, and the correlation parameter $\rho$ calculated as follows:

\begin{equation}\label{equ3}
\rho= \frac {<A'_{1} A'_{2}>}{(\sigma_{A'_{1}}\sigma_{A'_{2}})}.
\end{equation}

Here $A'_1$ and $A'_2$ represent the noise excursions with respect to the mean value of the corresponding amplitude and $\sigma_{A'_1}$ and $\sigma_{A'_2}$ are the sigma parameters obtained from the Gaussian fit of the amplitude distributions. 

\begin{figure}[!ht]
\begin{center}
\begin{tabular}{p{0.55\textwidth}@{\hspace{0.001\textwidth}}p{0.55\textwidth}}
\includegraphics[width=0.55\textwidth]{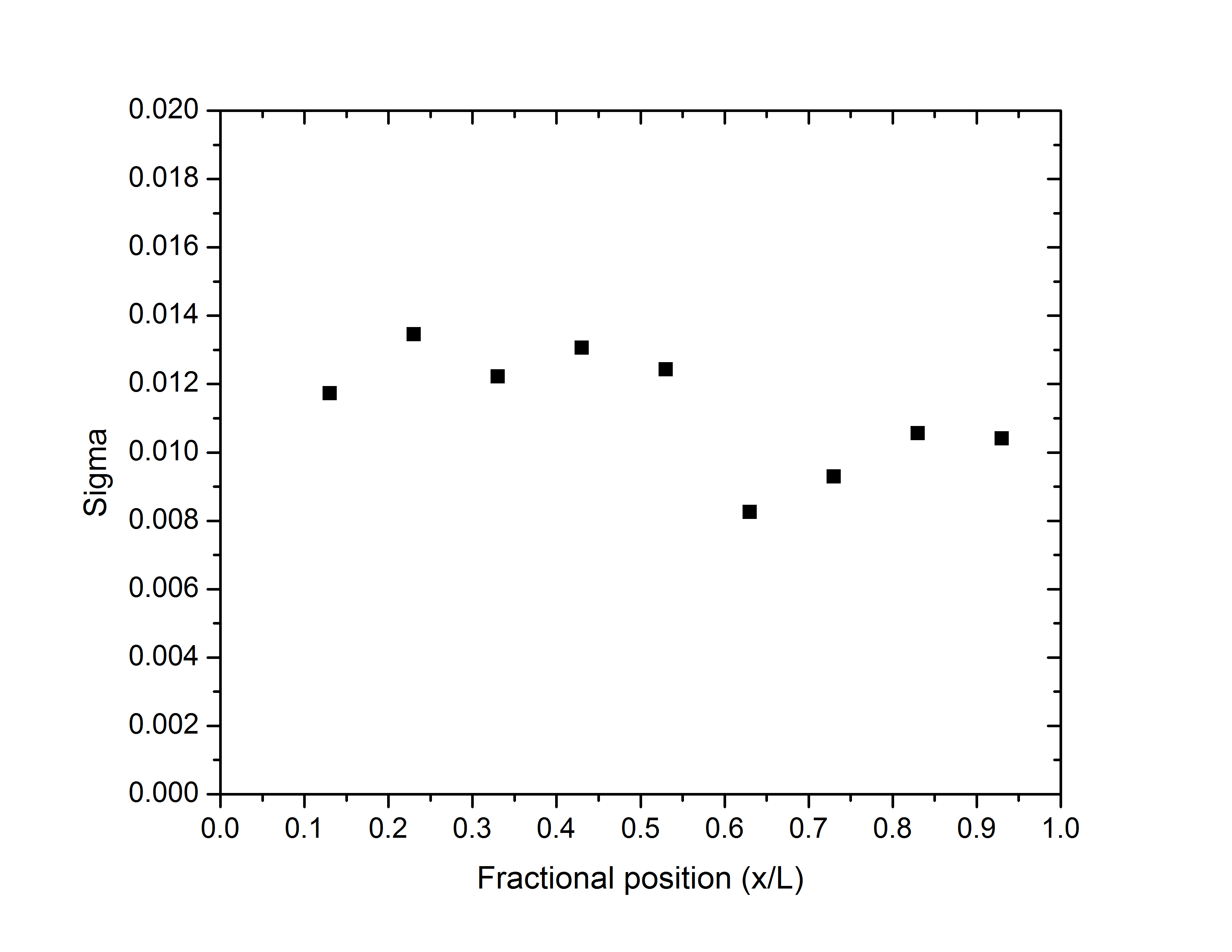} &
\includegraphics[width=0.55\textwidth]{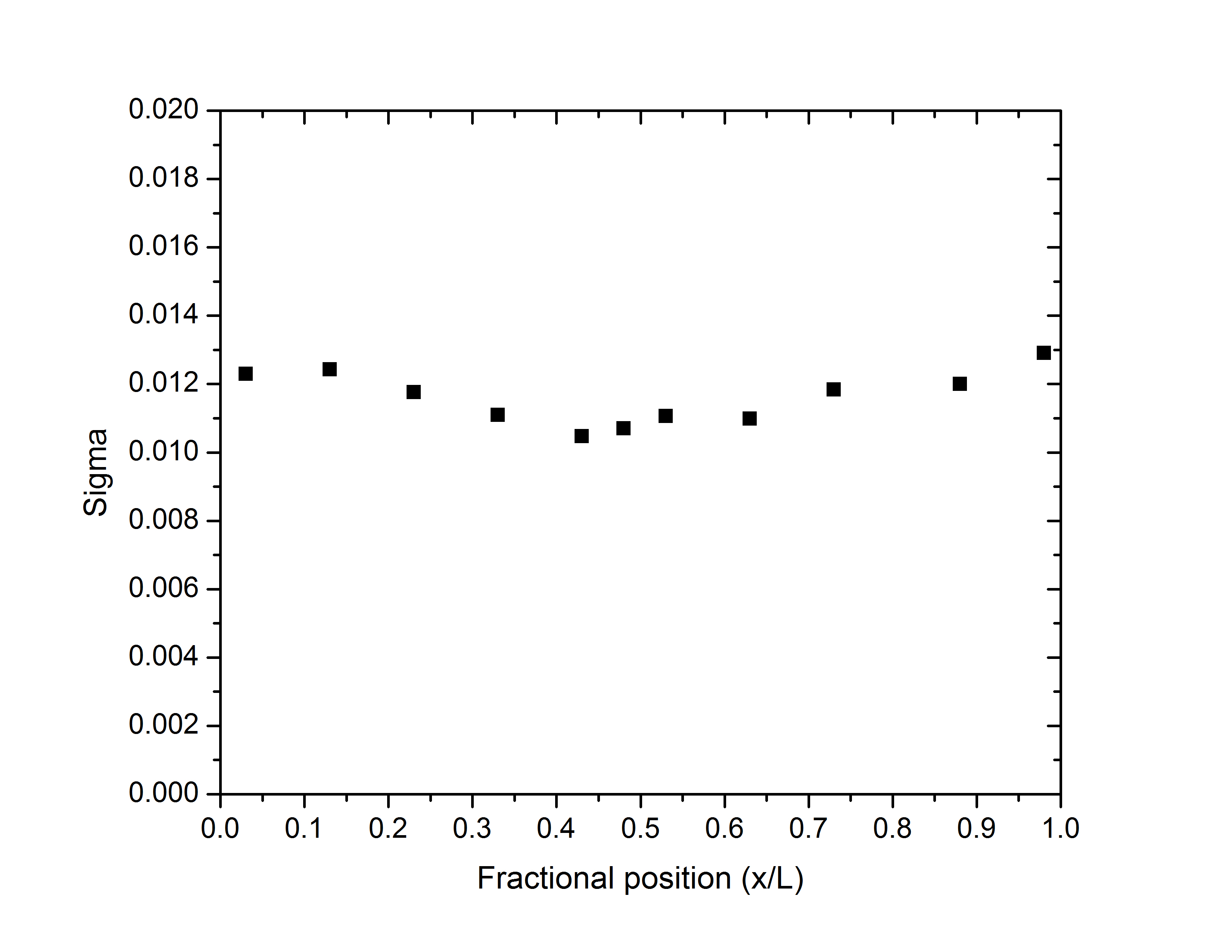} \\
\centerline{(a)}&\centerline{(b)}\\

\end{tabular}
\end{center}
\vspace{-0.7 cm}
\caption{\footnotesize{Calculated fractional position error for prototype with electrode resistance (a) $R\slash  l$ = 2.8~$\Omega \slash \mu$m  and  (b) $R\slash  l$ = 12.2~$\Omega \slash \mu$m.}}
\label{fig:sigma}
\end{figure}

The $\sigma$ value  is computed for each of the scan points. Figure \ref{fig:sigma} shows the error dependece on the position for both the sensors.
The $\sigma$ value strongly depends on the Signal to Noise ratio, that depends on the total charge created by the ionizing event as well as on its position along the strip.
The mean value of the spatial resolution obtained is 1.1\% and 1.2\% of  total strip length for the prototype with electrode resistance $R\slash  l$ = 2.8~$\Omega \slash \mu$m  and $R\slash  l$ = 12.2~$\Omega \slash \mu$m respectively.

\section{Conclusions}
We have introduced a novel 2D position-sensitive semiconductor detector concept based on the 
resistive charge-division readout method and manufactured prototypes using the standard technology of AC coupled
microstrip detectors. The implementation of resistive coupling electrodes allows to extract 
the information on the longitudinal coordinate of an ionizing event using the resistive 
charge-division method. 

Two proof-of-concept prototypes have been produced with strips 20~mm long and with different
linear resistances of the electrodes: $R\slash l$ = $2.8~\Omega \slash \mu$m and
$R\slash l$ = 12.2~$\Omega \slash \mu$m. A first investigation of their performance has been 
carried out using a Near Infra-Red laser and readout electronics based on the Beetle ASIC. 
Results show that the mean spatial resolution for a 6 MIP signal is 225~$\mu$m and 232~$\mu$m  for the two prototypes respectively.
 
An electrical simulation of the sensor equivalent circuit -including the amplifying 
and filtering stages- has been developed and benchmarked against the experimental data. 
The simulation allowed to highlight the effects of the propagation of the signal pulse along the dispersive electrodes: the amplitude attenuation and the increase of the peaking time that cause a systematic non constant ballistic deficit when non optimized front-end electronics is used to read the signal. The good agreement with the experimental results in reconstructing the fractional position of the signal generation point validates the electrical simulation as an adequate
tool for future sensor optimization.
    
This initial study demonstrates the feasibility of the resistive charge division method in a
fully fledged microstrip sensor with resistive electrodes. Specific studies on detection of minimum
ionizing particles are in progress to assess its soundness as tracking technology for the future
particle physics experiments; nevertheless, in its current conception, this implementation appears as a suitable
technology for highly ionizing particles as it is the case of neutron monitors based on conversion
layers or other nuclear imaging technologies ranging from Compton cameras to heavy-ion detection.

\acknowledgments
We thanks A.Candelori (INFN, Padova) for the clarifications concerning the SPICE model of ref.\cite{spice}; Gianluigi Casse (University of Liverpool) for the bonding of the sensors and
boards and Marko Dragicevic (HEPHY, Vienna) for contributing to the design of the mask of the sensors.

This work has been supported by the Spanish Ministry of Science and Innovation under grant
FPA2007-66387 and through the GICSERV program "Access to ICTS integrated nano-and micro
electronics cleanroom" of the same Ministry.


\begin{thebibliography}{12}
\bibitem{gas} H. Foeth, R. Hammarstrom, C. Rubbia, Nuclear Instruments and Methods in Physics Research 109 (1973) 521;

P. Schubelin, et al., Nuclear Instruments and Methods in physics Research 131 (1975) 39;

A. Feinberg, N. Horwitz, I. Linscott, G. Moneti, Nuclear Instruments and Methods in Physics Research 141 (1977) 277;

V. Radeka, P. Rehak, IEEE Transactions on Nuclear Science NS-26 (1979) 225;
\bibitem{pad} R.B. Owen, M.L. Awcock, IEEE Transactions on Nuclear Science NS-15 (1958) 290;
\bibitem{Radeka} V. Radeka, IEEE Transactions on Nuclear Science NS-21 (1974) 51;
\bibitem{board} J. K. Carman, et al., Nuclear Instruments and Methods in Physics Research A 646 (2011) 118;
\bibitem{Tur} R. Turchetta, Nuclear Instruments and Methods in Physics Research A 335 (1993) 44-58;
\bibitem{cnm} Centro Nacional de Microelectr\'onica, Campus Universidad Aut\'onoma de Barcelona. 08193 Bellaterra (Barcelona), Spain (http://www.imb-cnm.csic.es/);
\bibitem{TSVienna} T. Bergauer et al., Nuclear Instruments and Methods in Physics Research A 598 (2009) 86-88;
\bibitem{alibava} R.Marco-Hernandez and ALIBAVA collaboration,IEEE Transactions on Nuclear Science NS-56 (2009) 1642;
\bibitem{beetle} "Beetle - a readout chip for LHCb" http://www.kip.uni-heidelberg.de/lhcb/;
\bibitem{spice} N. Bacchetta et al. ,  IEEE Transactions on Nuclear Science NS-42 (1995) 459;
\bibitem{Spectre} Virtuoso Spectre data sheet, Cadence. Available at http://www.cadence.com/
products/custom ic/index.aspx.

 
\end{thebibliography}
\end{document}